\documentclass[
    10pt,
    twocolumn,
    letterpaper,
    aps,
    pra,
    superscriptaddress,
    showpacs,
    amsmath,
    floatfix,
]{revtex4-1}

\usepackage{graphicx}
\usepackage{color}
\usepackage{comment}

\usepackage{amssymb}
\usepackage{color}
\usepackage{graphicx}
\usepackage{braket}
\usepackage{here}
\usepackage[normalem]{ulem}
\begin{document}

\title{Nonlinear squeezing for measurement-based non-Gaussian operations in time domain}

\author{Shunya Konno}
\affiliation{Department of Applied Physics, School of Engineering,\\ The University of Tokyo, 7-3-1 Hongo, Bunkyo-ku, Tokyo 113-8656, Japan}
\author{Atsushi Sakaguchi}
\affiliation{Department of Applied Physics, School of Engineering,\\ The University of Tokyo, 7-3-1 Hongo, Bunkyo-ku, Tokyo 113-8656, Japan}
\author{Warit Asavanant}
\affiliation{Department of Applied Physics, School of Engineering,\\ The University of Tokyo, 7-3-1 Hongo, Bunkyo-ku, Tokyo 113-8656, Japan}
\author{Hisashi Ogawa}
\affiliation{Department of Applied Physics, School of Engineering,\\ The University of Tokyo, 7-3-1 Hongo, Bunkyo-ku, Tokyo 113-8656, Japan}
\author{Masaya Kobayashi}
\affiliation{Department of Applied Physics, School of Engineering,\\ The University of Tokyo, 7-3-1 Hongo, Bunkyo-ku, Tokyo 113-8656, Japan}
\author{Petr Marek}
\affiliation{Department of Optics, Palack\'y University, 17. listopadu 1192/12, 77146 Olomouc, Czech Republic}
\author{Radim Filip}
\affiliation{Department of Optics, Palack\'y University, 17. listopadu 1192/12, 77146 Olomouc, Czech Republic}
\author{Jun-ichi Yoshikawa}
\affiliation{Department of Applied Physics, School of Engineering,\\ The University of Tokyo, 7-3-1 Hongo, Bunkyo-ku, Tokyo 113-8656, Japan}
\author{Akira Furusawa}
\email{akiraf@ap.t.u-tokyo.ac.jp}
\affiliation{Department of Applied Physics, School of Engineering,\\ The University of Tokyo, 7-3-1 Hongo, Bunkyo-ku, Tokyo 113-8656, Japan}

\date{\today}

\begin{abstract}
Quantum non-Gaussian gate is a missing piece to the realization of continuous-variable universal quantum operations in the optical system. In a measurement-based implementation of the cubic phase gate, a lowest-order non-Gaussian gate, non-Gaussian ancillary states that has a property we call \emph{nonlinear squeezing} are required. This property, however, has never been experimentally verified. In this paper, we generate a superposition between a vacuum state and a single photon state whose nonlinear squeezing are maximized by the optimization of the superposition coefficients. The nonlinear squeezing is observed via real-time quadrature measurements, meaning that the generated states are compatible with the real-time feedforward and are suitable as the ancillary states for the cubic phase gate in time domain. Moreover, by increasing the number of the photons, it is expected that nonlinear squeezing can be further improved. The idea presented here can be readily extended to the higher-order phase gates [P.~Marek \textit{et al.}, Phys.~Rev.~A \textbf{97}, 022329 (2018)]. As such, this work presents an important step to extend the CV quantum information processing from Gaussian regime to non-Gaussian regime.
\end{abstract}

% 03.67.Lx Quantum computation architectures and implementations
% 42.50.Dv Quantum state engineering and measurements
% 42.50.Ex Optical implementations of quantum information
%          processing and transfer
% 42.65.-k Nonlinear optics

\maketitle

%update reference

%%%%%introduction%%%%%%%%%%%%%%%
\section{Introduction}

Continuous-variable (CV) quantum computation using optical system is currently one of the most promising approach to a scalable and practical quantum computation. As a recent progress, large-scale Gaussian cluster states, the computational resource states for measurement-based quantum computation \cite{PhysRevLett.86.5188,PhysRevLett.97.110501}, have been experimentally realized using time-domain-multiplexing method \cite{Yokoyama2013,doi:10.1063/1.4962732,Asavanant373,Larsen369}. By implementing appropriate measurements on the Gaussian cluster states, universal CV quantum operations can be realized. Even more recently, by combining basis-programmable homodyne measurements with time-domain cluster states, Gaussian operations, i.e.\ linear transformations of the quadrature operators, have been demonstrated \cite{2020arXiv200611537A,2020arXiv201014422L}. These experimental results demonstrate the potentials of the CV optical systems for quantum computation.

\begin{figure}
\centering
\includegraphics[width = \columnwidth]{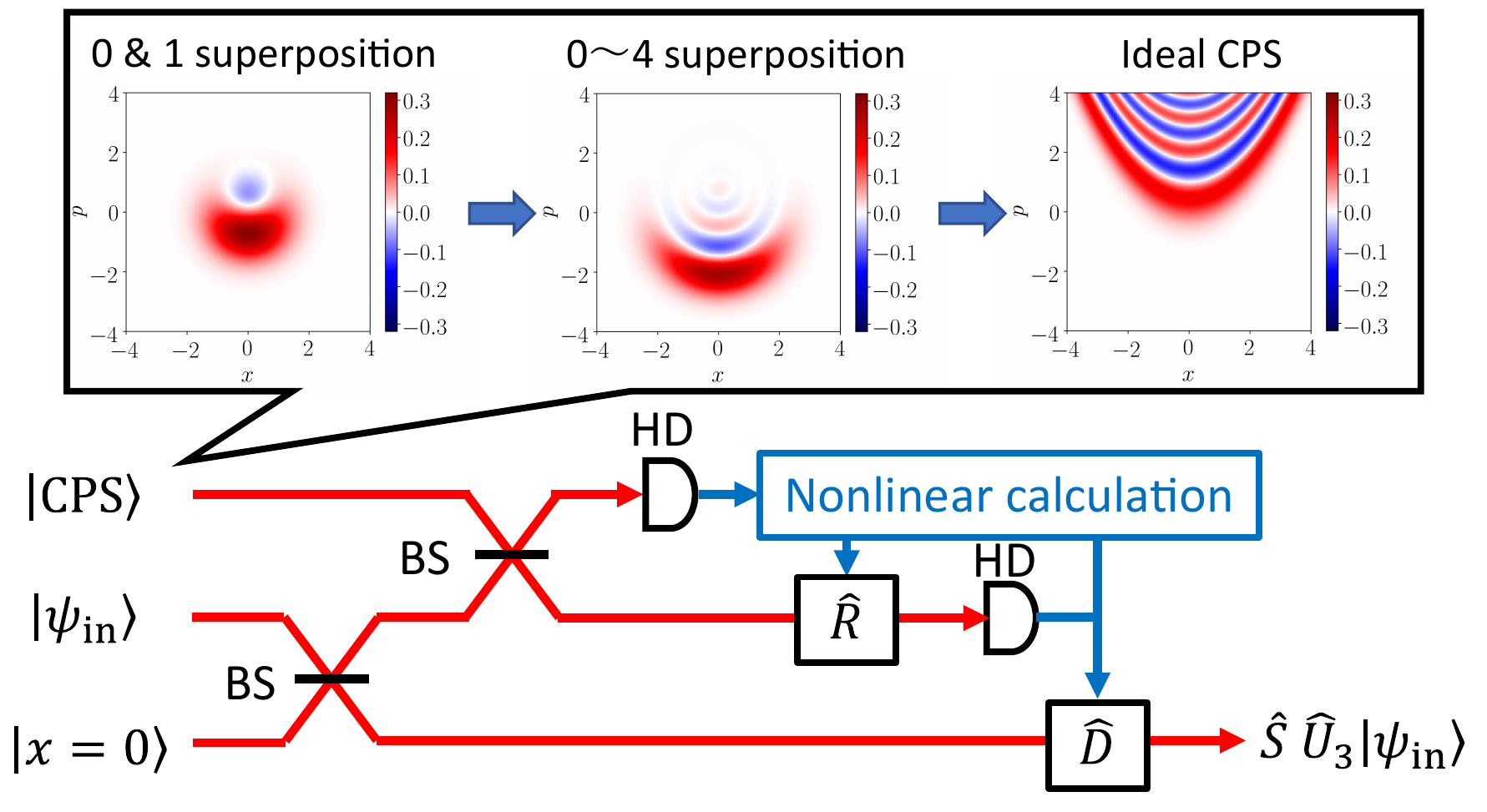}
\caption{Schematic diagram of implementation of cubic phase gate using non-Gaussian ancillary states, homodyne measurements, and nonlinear feedforwards. $\ket{\textrm{CPS}}$: cubic phase state; BS: beamsplitter; HD: homodyne detector; $\hat{R}$: phase rotation; $\hat{D}$: displacement operation; $\hat{S}$: squeezing operation; $\hat{U}_{3}$: cubic phase gate. The output of the above setup is a cubic phase gate followed by a squeezing operation. Because the ideal CPS is unphysical, we need to consider the approximations using the superpositions of Fock states. As the maximum photon number is increased, the nonlinear squeezing also approaches ideal limit. Note that we omitted displacement in $p$ direction of the ancillary states.}\label{fig:CPG_schematic}
\end{figure}

It is known, however, that in addition to the Gaussian operations, at least a single non-Gaussian operation is required to achieve the universality \cite{CVUniversality}. Moreover, because CV quantum computation that has no non-Gaussian elements can be effectively simulated with classical computer \cite{PhysRevLett.88.097904}, non-Gaussian element is a necessary requirement to achieve a useful quantum computation. One of the methods to realize non-Gaussian element in measurement-based quantum computation is by implementing nonlinear measurements, i.e. measurements that are nonlinear in quadrature basis \cite{PhysRevA.64.012310,PhysRevLett.97.110501}. Note that since homodyne measurements are linear in the quadrature operators, direct nonlinear measurements would require some sort of additional inline strong optical nonlinearity, which is very difficult to realize. Luckily, there exists a feasible alternative: nonlinear quadrature measurements can be implemented using ancillary states, homodyne measurements, and nonlinear feedforwards based on the measurement results \cite{GateTeleCV,PhysRevA.64.012310}. The implementations of the cubic phase gate (Fig.\ \ref{fig:CPG_schematic}) and its generalization, $N$-th-order phase gates, based on this approach have been proposed \cite{CPG2,NPG}. Some progresses and basic prototypes have already been made regarding the nonlinear feedforward in the time domain \cite{PhysRevA.90.060302,8426782}. Therefore, the next experimental hurdle that we must overcome is the generation of the non-Gaussian ancillary states.

As nonlinear feedforward operations introduce noises that are nonlinear in quadrature operators, the ideal ancillary states must be non-Gaussian states that are tailored to suppressing such noises. For example, an ancillary state of the cubic phase gate must suppress the noise with a form of $\hat{p}-3\kappa\hat{x}^2$, where $\kappa$ is a real number. This idea is similar to how the amount of the squeezing in the cluster states determines the amount of the noise added when they are used as resources for Gaussian operations \cite{PhysRevA.100.010301}. Therefore, we will call the suppression of the nonlinear noise in these non-Gaussian ancillary states as \emph{nonlinear squeezing} (NLSQ). In addition to high NLSQ, the ancillary states for nonlinear measurements should be compatible with the time-domain-multiplexing method. In time-domain CV quantum computation, modes are defined as localized temporal wave packets \cite{PhysRevA.83.062314}. Therefore, to obtain the quadrature values, we need to integrate the signals from the continuous homodyne measurements with the shape of the temporal modes. For general temporal modes, this integration introduces additional latency which can easily result in the decoherence of the system and has been one of the main technical difficulties of the measurement-based non-Gaussian gates that operate in the time domain. This difficulty can be overcome by engineering the shape of the wave packet of the ancillary states so that we can implement real-time quadrature measurements \cite{RealTime}. Experimentally, such real-time measurements have been demonstrated on the single photon states \cite{RealTime} and the cat states \cite{Asavanant:17}. However, generation of the ancillary states that possesses both NLSQ and the compatibility with the real-time measurement has not been achieved yet, as the NLSQ requires more optimization than the aforementioned quantum states. As a side note, this technical difficulty does not arise in the linear feedforward because it is commutable with the integrations with the temporal modes.

In this paper, we experimentally generate quantum states that exhibit NLSQ necessary for the implementation of the cubic phase gate \cite{CPG2}. In the ideal limit, this state is called cubic phase state and is infinitely nonlinearly squeezed. In general, achieving perfect NLSQ requires ancillary states that are unphysical, similar to how the generation of an ideal cluster state requires infinitely squeezed states which are also unphysical. As an alternative, we considered a superposition of Fock states that is truncated below a certain number of photon and optimize the NLSQ by selecting appropriate superposition coefficients \cite{CPG2}. Experimentally, we generate a superposition of a vacuum state and a single photon state, and show that there is indeed a parameter that maximize the NLSQ. The shape of the wave packets of the generated states are tailored so that we can implement the real-time measurements. The amount of the NLSQ observed here is higher than the amount that can be achieved with arbitrary Gaussian states, showing that the generated non-Gaussian states are indeed appropriate approximations as the ancillary states for cubic phase gate. Note that although there are already demonstration of generation of superpositions of Fock states up to three photons \cite{Yukawa:13}, to the best of our knowledge, this work is the first to show the NLSQ of the quantum states whose wave packet is compatible to the real-time quadrature measurements.

This paper is structured as follows. Sec.~\ref{sec:NLSQ_state} describes the concept of nonlinear squeezing and show the explicit example of the cubic phase gate which is relevant to the experimental results. Sec.~\ref{sec:method} explains the generation method of the superposition of the vacuum states and the single photon and how to tailor the temporal wave packet into an exponentially rising wave packet suitable for real-time measurements. Sec.~\ref{sec:setup} explains the experimental setup. The experimental results are discussed in Sec.~\ref{sec:results} which shows the generated states and their NLSQ. Finally, Sec.~\ref{sec:conclusion} gives a summary and future perspective.

%%%%%approximated ancilla%%%%%%%%%%%
\section{Nonlinear squeezing}
\label{sec:NLSQ_state}

We will denote $\hat{x}$ and $\hat{p}$ as the quadrature operators whose commutation relation is $[\hat{x},\hat{p}]=i$. We will consider a type of operation called $N$-th order phase gate which is given by a unitary operator 
\begin{equation}
\hat{U}_{N}(\kappa)=\exp\left(i\kappa\hat{x}^N\right),\label{eq:Nphasegate}
\end{equation}
where $\kappa$ is a parameter that determines the strength of the operations. Note that this operator becomes displacement operation when $N=1$ and squeezing operation when $N=2$ \cite{PhysRevA.90.060302}. When $N\geq3$, the above operation is a non-Gaussian operation which transforms Gaussian pure states to non-Gaussian ones. As $\hat{U}_{N}(\kappa)$ depends only on the operator $\hat{x}$, it leaves the operator $\hat{x}$ unchanged and transforms the operator $\hat{p}$ in the Heisenberg picture as
\begin{equation}
\hat{p}\to\hat{p}+N\kappa\hat{x}^{N-1}.
\end{equation}
In the implementation of $\hat{U}_{N}(\kappa)$ using the measurement-based approach, the ideal ancillary states $\ket{\psi_{\textrm{anc}}}$ are given by \cite{NPG}
\begin{equation}
\ket{\psi_\textrm{anc}}=\hat{U}_{N}(\kappa)\ket{p=0}.
\end{equation}
In addition to the representation of the ancillary states using the state vectors, the above ideal ancillary states are also the zero eigenstates of the following operators:
\begin{equation}
\mathrm{e}^{i \kappa \hat{x} ^N} \hat{p} \mathrm{e}^{-i \kappa \hat{x} ^N}=\hat{p} - N \kappa \hat{x}^{N-1}.\label{eq:NLquadrature}
\end{equation}
For any realistic ancillary states, the variances of the operator in Eq.\ \eqref{eq:NLquadrature} will be nonzero. However, the smaller this variance is the smaller the contribution of the ancillary state to the noises in the $N$-th order phase gate becomes. As a concrete example, let us consider the implementation of the cubic phase gate in Fig.\ \ref{fig:CPG_schematic}. The Heisenberg picture input--output relation of this circuit is given by \cite{CPG2} 
\begin{align}
\hat{x}_\textrm{out}=&\frac{1}{\sqrt{2}}\hat{x}_\textrm{in}-\frac{1}{\sqrt{2}}\hat{x}_\textrm{sqz}\\
\begin{split}
\hat{p}_\textrm{out}=&\sqrt{2}\left(\hat{p}_\textrm{in}+\frac{3\kappa}{2\sqrt{2}}\hat{x}_\textrm{in}^2\right)+(\hat{p}_\textrm{anc}-3\kappa\hat{x}_\textrm{anc}^2)\\
&+3\gamma\left(\hat{x}_\textrm{in}\hat{x}_\textrm{sqz}+\frac{1}{2}\hat{x}_\textrm{sqz}^2\right).
\end{split}\label{eq:pcubicout}
\end{align}
The first terms on the right show the ideal cubic phase gate operations (with additional squeezing operations), while the second term of $\hat{p}_\textrm{out}$ is the nonlinear noise term due to the ancillary states. We observe that on the right hand side of Eq.\ \eqref{eq:pcubicout}, the signs in front of $\hat{x}_\textrm{in}^2$ and $\hat{x}_\textrm{anc}^2$ are opposite. This is due to the difference between the active operations and the resource state needed for it. The amount of the NLSQ in the ancillary states for the cubic phase gate will corresponds to the noise contributions of this second term. On the other hand, the noises due to the terms concerning $\hat{x}_\textrm{sqz}$ approach zero as the squeezing level of the ancillary squeezed state increases.

To define the NLSQ in an operational way that is experimentally useful, let us consider an approximation of the ancillary state $\ket{\psi_\textrm{NL}}$ given by
\begin{equation}
\ket{\psi_\mathrm{NL}} = \hat{S}(\lambda) \sum_{k = 0}^M c_k \ket{k},\label{eq:psiNL}
\end{equation}
which is a superposition of Fock states up to $M$ photons, followed by the squeezing operations $\hat{S}(\lambda)$. The reason we consider the squeezing operation here is as follows. If we implement squeezing operation and its Hermitian conjugate before and after the $\hat{U}_{N}(\kappa)$, we can easily show that
\begin{align}
\hat{S}(\lambda)\hat{U}_{N}(\kappa)\hat{S}^\dagger(\lambda)=\hat{U}_{N}(\kappa/\lambda^N),
\end{align}
where we assume that $\hat{S}^\dagger(\lambda)\hat{x}\hat{S}(\lambda)=\lambda\hat{x}$ and $\hat{S}^\dagger(\lambda)\hat{p}\hat{S}(\lambda)=\hat{p}/\lambda$. As the squeezing operation is equivalent to simply adjustment of the strength parameter $\kappa$, the optimal superposition coefficients $\{c_{k}\}$ are identical for all $\kappa$ and the optimal ancillary states of the different $\kappa$ are related simply via squeezing operations. This squeezing operation can be either implemented actively \cite{MIQC1,PhysRevLett.113.013601} or can be compensated by adjusting the splitting ratios of the beamsplitters in the circuit of Fig.\ \ref{fig:CPG_schematic}. Therefore, without loss of generality, we will assume that $\kappa=1$ and consider the cubic phase gate, i.e. the case where $N=3$ in Eq.\ \eqref{eq:Nphasegate} which is the case of our main interest. Then, we can define optimal nonlinear variance for any quantum state $\hat{\rho}$ as the minimum
\begin{equation}
V_{\hat{\rho}}^{\mathrm{opt}} = \min_{\lambda > 0} \left\{ \mathrm{Tr} \left[ \left( \Delta \hat{y}\right)^2 \hat{\rho} \right] \right\}, \label{eq:NLvariance}
\end{equation}
where $\hat y = \lambda \hat p -3\left(\frac{\hat x}{\lambda}\right)^2$, $\Delta\hat{y}=\hat{y}-\mathrm{Tr}(\hat{y}\hat{\rho})$, and $\mathrm{Tr}(\cdot)$ is a trace of an operator. This optimization with respect to $\lambda>0$ guarantees that the optimal variances are the same for all the states that are related to each other via squeezing operations. We consider only positive $\lambda$ here as this preserve the form of the variances in Eq.\ \eqref{eq:NLquadrature}, whereas the negative $\lambda$ would represent squeezing and phase shift, but it would not lead to squeezed variance for the same state. To quantify how much a quantum state $\hat{\rho}$ is nonlinearly squeezed, we compare the amount of $\hat{V}_{\hat{\rho}}^{\mathrm{opt}}$ to the smallest amount achievable with Gaussian states. Since Eq.\ \eqref{eq:NLvariance} is invariant with respect to sign flip of $\hat{x}$, the optimal Gaussian state should be symmetric with respect to the $p$ axis. It can be also easily checked that the optimal displacement of $\hat{x}$ is zero and that displacement of $\hat{p}$ has no effect. On the other hand, the effect of the squeezing is considered in the optimization with respect to $\lambda$. As a result, the optimal Gaussian state is the squeezed vacuum state and the Gaussian limit is given by $V_{\ket{0}\bra{0}}^\mathrm{opt}$ because the squeezing is included in the minimization over lambda. The quantum state is therefore nonlinearly squeezed when 
\begin{equation}
\frac{V_{\hat{\rho}}^\mathrm{opt}}{V_{\ket{0}\bra{0}}^\mathrm{opt} } < 1 
\end{equation}
and the amount of NLSQ is directly given by this ratio. For any state to be useful as an ancillary state, it must exhibit NLSQ as the vacuum state has a better noise performance as an ancillary state if NLSQ is not present. See appendix A for more detailed calculations.

The simplest quantum state capable of exhibiting the nonlinear squeezing can be prepared as a superposition of Fock states $c_0 \ket{0} + c_1 \ket{1}$. The parameters $c_0$ and $c_1$ can be optimized to obtain maximal relative reduction of variance $V_{\hat{\rho}}^{\mathrm{opt}}/V_{\ket{0}\bra{0}}^{\mathrm{opt}}$, which, for pure states, is obtained in state $0.79\ket{0} - 0.61i\ket{1}$ \cite{CPG2}. However, in the realistic scenario with losses, the optimal state can be different. 

%%%%%generation method%%%%%%%%%%%
\section{Generation method}
\label{sec:method}
\begin{figure}[htp]
\centering
\includegraphics[width = \columnwidth]{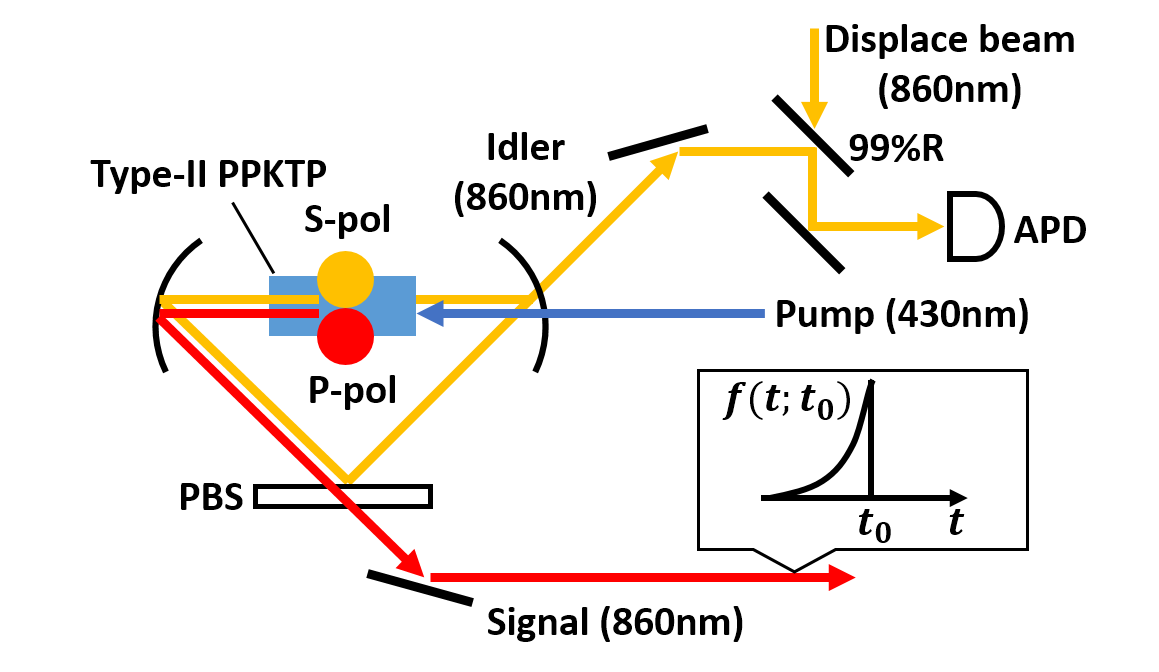}
\caption{Schematic diagram of the asymmetric optical parametric oscillator (OPO) based on type-II PPKTP used in this experiment. PBS: polarization beamsplitter, APD: avalanche photodiode. The displace beam is used in the preparation of the superposition between $\ket{0}$ and $\ket{1}$.}\label{fig:AOPO}
\end{figure}
To generate the superposition of Fock states with an exponentially rising wave packet, we use an asymmetric optical parametric oscillator (OPO) \cite{RealTime} which is shown in Fig.\ \ref{fig:AOPO}. A superposition of the vacuum state and the single photon state is generated by a heralding scheme. Initially, a two-mode squeezed state  
\begin{equation}
\ket \psi = \sqrt{1-\vert q\vert^2}\left(\ket{0}_\text{s}\ket{0}_\text{i}+q\ket{1}_\text{s}\ket{1}_\text{i}+\mathcal O(\vert q\vert^2)\right)
\end{equation}
is generated from an OPO where $q$ is a parameter representing the strength of the two-mode squeezing, which we assume to be sufficiently small. \lq{}s\rq{} indicates signal mode and \lq{}i\rq{} indicates idler mode. These two modes are separated and we implement a weak displacement operation on the idler mode which results in
\begin{align}
\begin{split}
\hat D_{\mathrm{i}}(\alpha)\ket \psi=&\exp(\alpha\hat{a}_\text{i}^\dagger-\alpha^\ast\hat{a}_{\text{i}})\ket{\psi}\\%\sim (1+\alpha \hat a^\dag_\text{i}-\alpha^*\hat a_\text{i})\ket \psi\\
\sim&(\ket{0}_\mathrm{s} - \alpha^{*} q \ket{1}_\mathrm{s} )\ket{0}_\mathrm{i}+ (\alpha\ket{0}_\text{s}+q\ket{1}_\text{s}) \ket{1}_\mathrm{i}\\
&+\mathcal{O}\left((\vert q\vert+\vert\alpha\vert)^2\right)
\end{split}
\label{eq:displacement}
\end{align}
where $\alpha$ represents the amount of the displacement and we assume $|\alpha| \ll 1$. Then, we do the photon detection of the idler mode. When a photon number state $\ket{1}$ is detected, a superposition between $\ket{0}$ and $\ket{1}$ is induced in a signal mode:
\begin{align}
_\text{i}\bra{1}\hat D_{\mathrm{i}}(\alpha)\ket \psi \sim \alpha\ket{0}_\text{s}+q\ket{1}_\text{s}.
\end{align}
The superposition coefficients are experimentally adjusted by changing the parameter $\alpha$ of the displacement operation and the strength $q$ of the two-mode squeezing.

In the asymmetric OPO used in this experiment, the signal mode and the idler mode have different polarization and one of the mirrors of the OPO is a polarization beamsplitter. This OPO is designed so that the signal mode is nonresonant to the OPO. Ideally, the wave packet of the signal mode when the photon is detected at the idler mode is an exponentially rising shape \cite{RealTime},
\begin{equation}
f(t;t_0) \propto \mathrm{e}^{-\gamma \vert t-t_0\vert/2}\Theta(t_0-t),
\end{equation}
where $\gamma$ is a parameter determined by the bandwidth of the OPO and $\Theta(\cdot)$ is a step function.

In the actual experiment, there are additional filtering cavities in idler mode which are used to remove unwanted frequency modes and they function as additional Lorenzian filters in the frequency domain. For the case where there are two filtering cavities, the shape of the wave packet when these filters are taken into an account can be expressed as follows \cite{RealTime}
\begin{equation}
f(t;t_0) \propto \sum_{n=1}^3c_ne^{-\gamma_n\vert t-t_0\vert/2}\Theta(t_0-t),\label{eq:3rdLPF}
\end{equation}
where $\gamma_\text 1$, $\gamma_2$, $\gamma_3$ correspond to the bandwidth of each cavity (OPO and the two filtering cavities), $c_1 = 1/(\gamma_2-\gamma_1)(\gamma_3-\gamma_1)$, and $c_2$ and $c_3$ are its cyclic permutations.

%%%%%experimental setup%%%%%%%%%%%
\section{Experimental setup}
\label{sec:setup}

%%%
\begin{figure*}[th]
\centering
\includegraphics[width = 180mm]{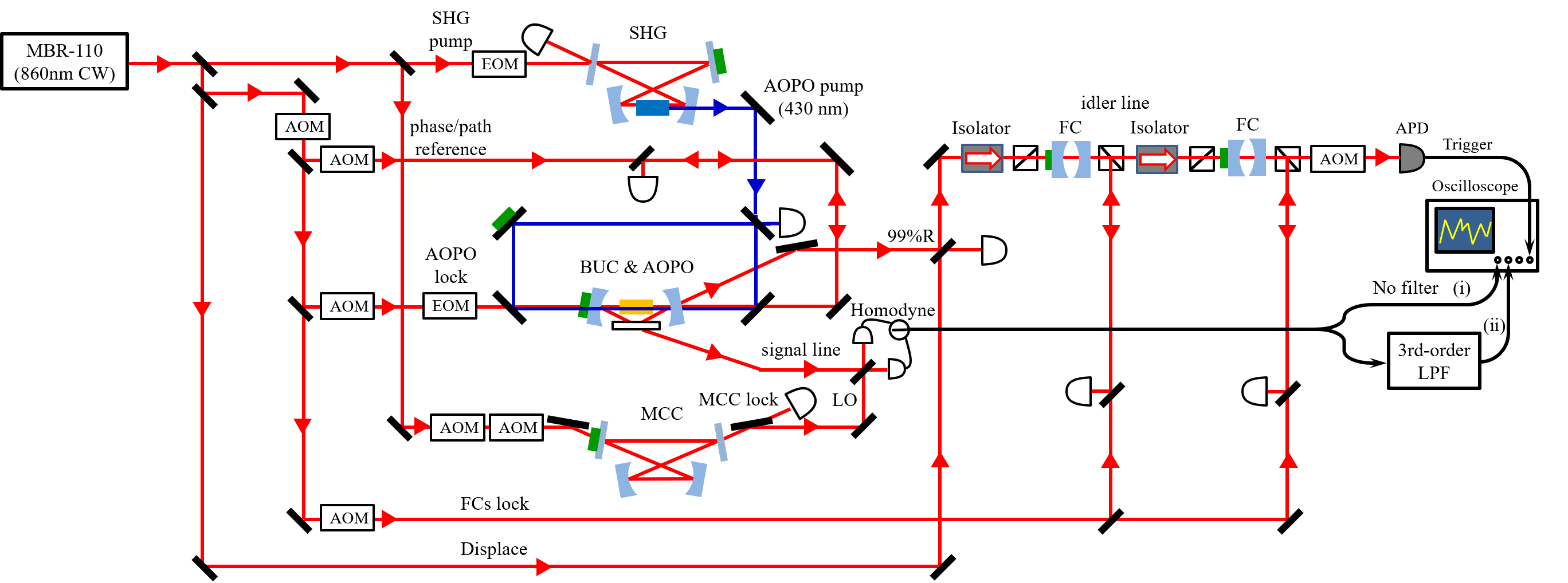}
\caption{Experimental setup for the generation of the superposition between $\ket{0}$ and $\ket{1}$. SHG: second harmonic generator. EOM: electro-optic modulator. AOM: acousto-optic modulator. AOPO: asymmetric optical parametric oscillator. BUC: buildup cavity. MCC: mode cleaning cavity. FC: filter cavity. 99\%R: 99\% reflectivity. APD: avalanche photodiode. LO: local oscillator beam. LPF: low-pass filter. }\label{fig:ExperimentalSetup}
\end{figure*}
%%%
Figure \ref{fig:ExperimentalSetup} shows the experimental setup. A continuous wave (CW) titanium sapphire laser (MBR-110) operating at a wavelength of 860 nm is used as a light source of the experiment. A part of the laser beam is sent to a bow-tie shaped second harmonic generator and is converted to a 430~nm CW pump beam for the asymmetric OPO. In order to enhance the pump power and stabilize the optical path of the pump beam, a buildup cavity is constructed around the OPO. The asymmetric OPO is a triangle-shaped cavity with a linewidth of 33.7 MHz at the half-width half-maximum (HWHM) and a periodically poled KTiOPO$_4$ crystal with type-II phase matching is placed inside the cavity. One of the mirrors of the OPO is a plate polarization beamsplitter (PBS). The s-polarized idler mode is reflected by this PBS, thus it is resonant to the OPO , while the p-polarized signal mode passes through the PBS and is off-resonant to the OPO. This results in the exponentially-rising wave packet explained in Sec.\ \ref{sec:method}.

The displacement operation is implemented on the idler mode by interfering it with a coherent beam (displace beam) at a beamsplitter with 99\% reflectivity. The absolute value of the parameter $\alpha$ in Eq.\ \eqref{eq:displacement} is adjusted by changing the power of this coherent beam and the phase of $\alpha$ is adjusted by the interference phase at this beamsplitter. After the displacement operation, there are two filter cavities which remove unwanted frequency sideband modes. Both filtering cavities are Fabry-Perot cavities whose linewidths at the HWHM are 140.1~MHz and 90.9~MHz, respectively. The idler photons are then detected by an avalanche photodiode (APD) and the electrical signals from the APD are sent to the oscilloscope and used as the measurement triggers 

%chopping
To stabilize the optical cavities and the relative phases at each beamsplitter, lock beams and phase reference beams are employed. During the state generations, however, as these beams might enter the APD and result in fake triggers, they are switched off using the acousto-optic modulators (AOMs). When these beams are turned off, the feedback controls are also off and the voltages of each feedback component are kept at the state right before. This method is called sample \& hold and is widely used in the generation of the non-Gaussian states \cite{RealTime,Yukawa:13,Asavanant:17,Wakui:07} .

To characterize the generated states, we perform homodyne measurements on the signal mode. A mode-cleaning cavity is used to spatially filter the local oscillator (LO) beam into a TEM\textsubscript{00} mode and the power of the LO beam is set to $\sim10$~mW. The homodyne detector has a flat frequency response up until $\sim200$~MHz, which is much broader than the linewidth of the asymmetric OPO, meaning that the effects of the finite bandwidth of the homodyne detector are negligible.

The electrical signals from the homodyne detector is recorded by the oscilloscope. The sampling rate of the oscilloscope is 5GS/s and the width of each frame is about 200~ns, centered around the trigger signal of the APD. The electrical signals from the homodyne detector is split into two paths. While one of them is connected directly to the oscilloscope [signal (i) in Fig.~\ref{fig:ExperimentalSetup}] and is used in the digital post-process method, the other is filtered by a third-order lowpass filter (LPF) whose time response function is designed to match the shape of the wave packet [signal (ii) in Fig.~\ref{fig:ExperimentalSetup}]. The electrically filtered signals are then recorded by the oscilloscope and these signals correspond to the quadrature values in the real-time measurement. 

For each generated state, quadrature in six measurement bases (0$^\circ$, 30$^\circ$, 60$^\circ$, 90$^\circ$, 120$^\circ$, and 150$^\circ$ from the axis of the $\hat{x}$ quadrature) with about 21,000 events for each basis are measured. We reconstruct the density matrices of the generated states using the maximum likelihood method \cite{Lvovsky_2004} from the obtained quadrature values of both the digital postprocessing method and the real-time measurement method. The NLSQ of each generated state is then calculated from the reconstructed density matrix.

%%%%%experimental results%%%%%%%%%%
\section{Experimental results}
\label{sec:results}
\begin{figure}[h]
\centering
\includegraphics[scale = 0.37]{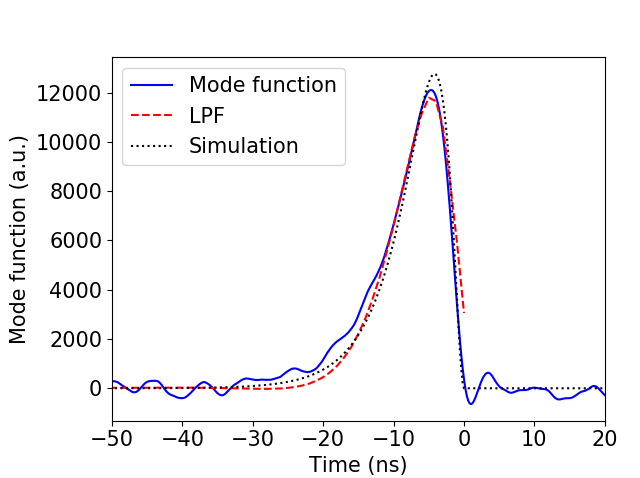}
\caption{Temporal mode function of the generated states. Blue solid curve: estimated temporal mode. Black dotted curve: temporal mode calculated by substituting the bandwidths of the cavities into Eq.~\eqref{eq:3rdLPF}. Red dashed curve: weighting function of the third-order LPF for the real-time measurement.}\label{fig:Modefunction}
\end{figure}

We generate superposition of 0 and 1 photon states with 9 different superposition coefficients. Note that the displacement beam is turned off when we generate the single photon state. We use the single photon states to first estimate the shape of the temporal mode function of the wave packet of the generated states. The estimation is done by using the principle component analysis on the homodyne signals without the LPF [signal (i) in Fig.~\ref{fig:ExperimentalSetup}] and the results are shown in Fig.~\ref{fig:Modefunction}. The estimated temporal mode function clearly exhibits the exponentially-rising feature and its mode matching to the theoretically predicted mode is about 98 \%. We also design a LPF whose time response function match with the temporal mode function. The time response function of the LPF is a red dashed curve in Fig.~\ref{fig:Modefunction} and its mode matching to the temporal mode of the generated state is about 97\% which is sufficient for the real-time measurements.

%%%
\begin{figure*}
\centering
\includegraphics[width = \textwidth]{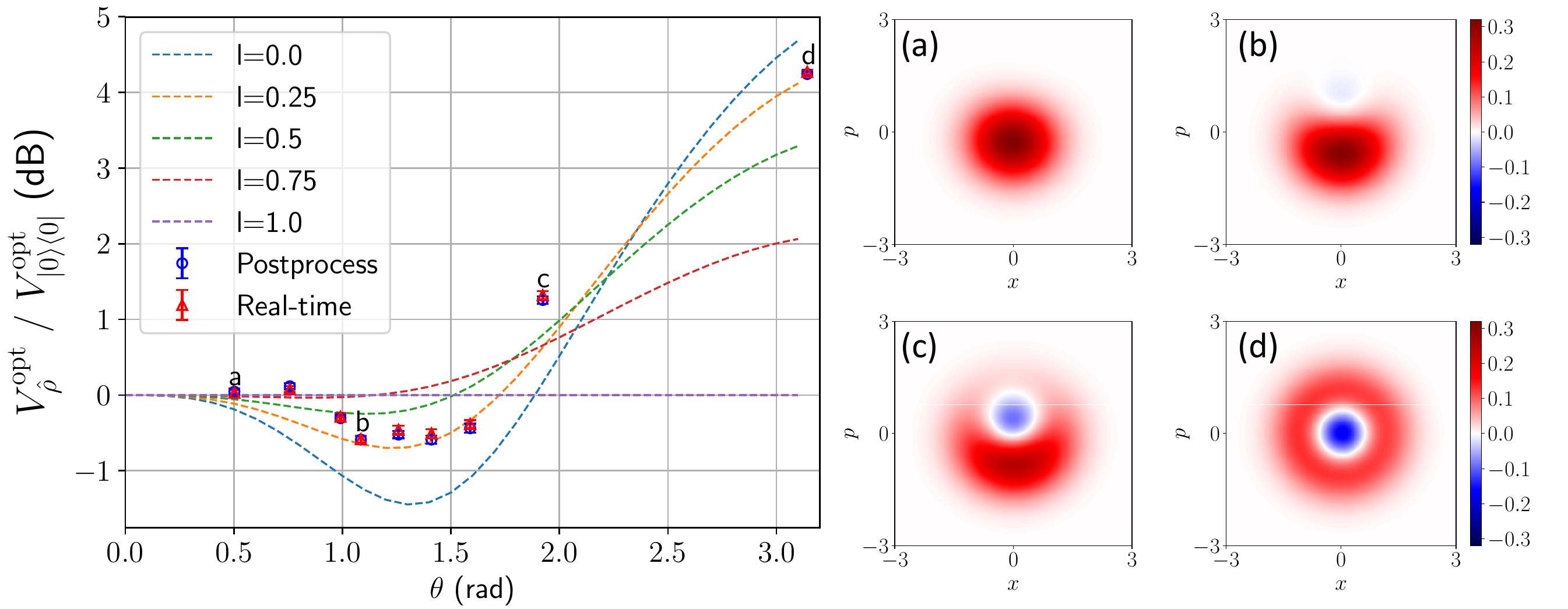}
\caption{NLSQ and some representative Wigner functions of the generated states. For the subfigure on the left, the blue circle markers and the red triangle markers correspond to the NLSQ of the states calculated by using the digital postprocessing method and the real-time measurement method, respectively. The dashed lines are theoretical predictions when there are optical losses of $L$. The subfigure (a) to (d) on the rights are the Wigner functions of the generated states that corresponded to each point of the plots on the left. Note that these are the Wigner functions for the real-time measurement method.}\label{fig:exp_result}
\end{figure*}
%%%
Figure \ref{fig:exp_result} shows the values of the nonlinear squeezing of the generated superpositions states and Wigner functions of some of the representative states. Note that the Wigner functions shown here are based on the real-time measurements. In general, for the case of the pure state, the superposition of $\ket{0}$ and $\ket{1}$ can be parametrized as
\begin{align}
\ket{\psi(\theta,\phi)}=\cos\left(\frac{\theta}{2}\right)\ket{0}+e^{i\phi}\sin\left(\frac{\theta}{2}\right)\ket{1}.
\end{align}
Although the actual generated states are not pure state, the parameter $\theta$ is adjusted by looking at the ratio of the count rate at the APD when the displacement beam is presented to when it is not presented, which corresponds to $(1+\tan^2(\theta/2)):\tan^2(\theta/2)$ when the state is a pure state. The parameter $\theta$ in the Fig.\ \ref{fig:exp_result} is thus the parameter we adjust. On the other hand, the parameter $\phi$ is set to $\phi=3\pi/2$ by locking the reference phase of the beamsplitter at the displacement operation.

From Fig.\ \ref{fig:exp_result} we can see that not all parameters result in the superposition with NLSQ. One of the interesting feature of the NLSQ is that although the degree of the non-Gaussianity of the quantum state is usually associated with how much the negative region the Wigner function has \cite{Kenfack_2004}, having large negative regions does not necessarily result in large NLSQ. For example, we can see that the single photon state, a state with most negativity in our consideration, actually performs worst, with the NLSQ of $4.24\pm0.03$ dB and $4.27\pm0.03$ dB for the digital postprocessing and the real-time method, respectively. Therefore, in terms of noises of the cubic phase gate, vacuum state performs better than the single photon state. On the other hand, we observe clear NLSQ for the region where the parameter $\theta$ is between 1.0 to 1.6. For example, at $\theta=1.09$ rad, which corresponds to point (b) in Fig. \ref{fig:exp_result}, the NLSQ are $-0.59\pm0.04$ dB and $-0.58\pm0.04$ dB for the digital postprocessing method and the real-time method, respectively. These values are better than the values that can be achieved with arbitrary Gaussian states (see also Appendix A). To also check whether the values of the NLSQ of the generated states matched with the experimental parameters or not, we also plotted theoretical predictions for various optical loss $L$. From the plot, we can see that the experimental values of NLSQ can be relatively well-explained by the $L=0.25$ which is on the same order with the experimental parameter. For further detailed discussions regarding the generated states, see Appendix B and C. These results indicate that by making a non-Gaussian superposition, we can experimentally generate states with NLSQ that also have an exponentially-rising wave packet appropriate for the implementation of the cubic phase gate in the time domain.

%%%%%experimental results%%%%%%%%%%
\section{summary and outlook}
\label{sec:conclusion}

In summary, we have generated superposition between vacuum state and single photon state with superposition coefficients such that it exhibits the NLSQ required for the cubic phase gate. Moreover, the generated states are in an exponentially-rising wave packet, which let us use real-time quadrature measurements to evaluate the state. Therefore, the generated ancillary states are compatible to the currently promising time-domain-multiplexing method of the optical CV quantum computation. Although this experiment is the first to evaluate the NLSQ and the values are still relatively low, by increasing the number of the photon and improve the optical losses of the system, we can expect that the NLSQ can be much improved. Moreover, this idea of NLSQ is not limited to the cubic phase gate, but can be applied to higher-order phase gate as well \cite{NPG}. Therefore, this work is a crucial step to extend the boundary of the optical CV quantum information processing from the already widely researched Gaussian regime into the non-Gaussian regime.

%%%%%acknowledgement%%%%%%%%%%%
\section*{Acknowledgments}
This work was partly supported by JSPS KAKENHI (Grant No. 18H05207, No. 18H01149, and No. 20K15187), the Core Research for Evolutional Science and Technology (CREST) (Grant No. JPMJCR15N5) of the Japan Science and Technology Agency (JST), UTokyo Foundation, and donations from Nichia Corporation. P.~M.~and R.~F.~acknowledge grant GA18-21285S of the Czech Science Foundation and also support by national funding from MEYS and European Union’s Horizon 2020 (2014–2020) research and innovation framework programme under grant agreement No. 731473 (project 8C20002 ShoQC). Project ShoQC has received funding from the QuantERA ERA-NET Cofund in Quantum Technologies implemented within the European Unions Horizon 2020 Programme.

%%%%%Appemdix A%%%%%%%%%%%
\section*{APPENDIX A: PARAMETERLESS NONLINEAR SQUEEZING}
\label{sec:appendixA}
 Nonlinear squeezing of a quantum state can be fully characterized by the ratio $V_{\hat{\rho}}^{\mathrm{opt}}/V_{\ket{0}\bra{0}}^{\mathrm{opt}}$, where $V_{\hat{\rho}}^{\mathrm{opt}} = \min_{\lambda > 0} \{ \mathrm{var}_{\hat{\rho}}\hat{y}\} $with $\hat y = \lambda \hat p -3\left(\frac{\hat x}{\lambda}\right)^2$ and the variance is taken in state $\hat{\rho}$. Do note that restricting $\lambda>0$ does not matter for the implementation of the quantum gate as it can be compensated by squeezing operations and the beamsplitter. To prove this assertion, let us first show $V_{\ket{0}\bra{0}}^{\mathrm{opt}}$ is the minimal variance achievable for any Gaussian state:
\begin{equation}
 V_\mathrm{Gauss}^\mathrm{opt} = \min_{\hat{\rho} \in \mathcal{G}} \min_{\lambda>0} \mathrm{Tr} \left[ \left( \Delta \hat{y}\right)^2 \hat{\rho} \right]
\end{equation}
This follows from the nature of operator $\hat{y}$. Since it is symmetric with respect to $\hat{x}$, the optimal Gaussian state will be, up to a displacement that can be neglected, a squeezed vacuum state with quadrature operators $\hat{p}_g = g\hat{p}_0$ and $\hat{x}_g = \hat{x}_0/g$, where $\hat{x}_0$ and $\hat{p}_0$ are quadrature operators of vacuum state. The minimal variance for a Gaussian state can then be rewritten as 
\begin{equation}
\min_{g>0} \min_{\lambda > 0} \mathrm{Tr} \left\{ \left[ \Delta (g \lambda \hat{p}_0 - 3 \frac{\hat{x}_0^2}{g^2\lambda^2}) \right]^2 \hat{\rho} \right\} ,
\end{equation}
and it can be immediately seen that $V_\mathrm{Gauss}^\mathrm{opt} = V_{\ket{0}\bra{0}}^\mathrm{opt}$. The optimal Gaussian state is the vacuum state due to the flexibility in $\lambda>0$.

We can now define, for any single value of nonlinear parameter $\kappa$, the nonlinear variance as 
\begin{equation}
V_{\hat{\rho}}(\kappa) = \min_{\lambda>0} \mathrm{var}_{\hat{\rho}} (\lambda \hat{p} - \kappa \frac{\hat{x}^2}{\lambda^2})
\end{equation}
and the relative nonlinear variance as 
\begin{equation}
R(\kappa) = \frac{V_{\hat{\rho}}(\kappa)}{V_{\ket{0}\bra{0}}^\mathrm{opt}(\kappa)}
\end{equation}
By taking a different nonlinear parameter $\kappa' = u^3 \kappa$, we arrive at
\begin{align}
V_{\hat{\rho}}(\kappa') &=  \min_{\lambda>0} \mathrm{var}_{\hat{\rho}} (\lambda \hat{p} - \kappa u^3 \frac{\hat{x}^2}{\lambda^2}) \nonumber  \\
&= u^2 \min_{\lambda>0} \mathrm{var}_{\hat{\rho}} (\frac{\lambda}{u} \hat{p} - \kappa \frac{u^2}{\lambda^2} \hat{x}^2) \nonumber  \\
&= u^2 V_{\hat{\rho}}(\kappa).
\end{align}
We can now see that $R(\kappa)$ does not actually depend on $\kappa$ and we can therefore choose it as we see fit.

%%%%%%%Appendix B%%%%%%%%
\section*{APPENDIX B: Correlations between digital postprocessing and real-time measurements}
\label{sec:appendixB}
In the main text, we have shown that the values of the NLSQ are similar for both the case when the states are evaluated using digital postprocessing and the real-time measurements. However, to evaluate whether we have succeeded in the real-time measurements or not, we have to show that the quadrature values that are obtained via the real-time method are the same with that of the postprocessing method for every single homodyne measurement.
\begin{figure}[htp]
\centering
\includegraphics[width=0.8\columnwidth]{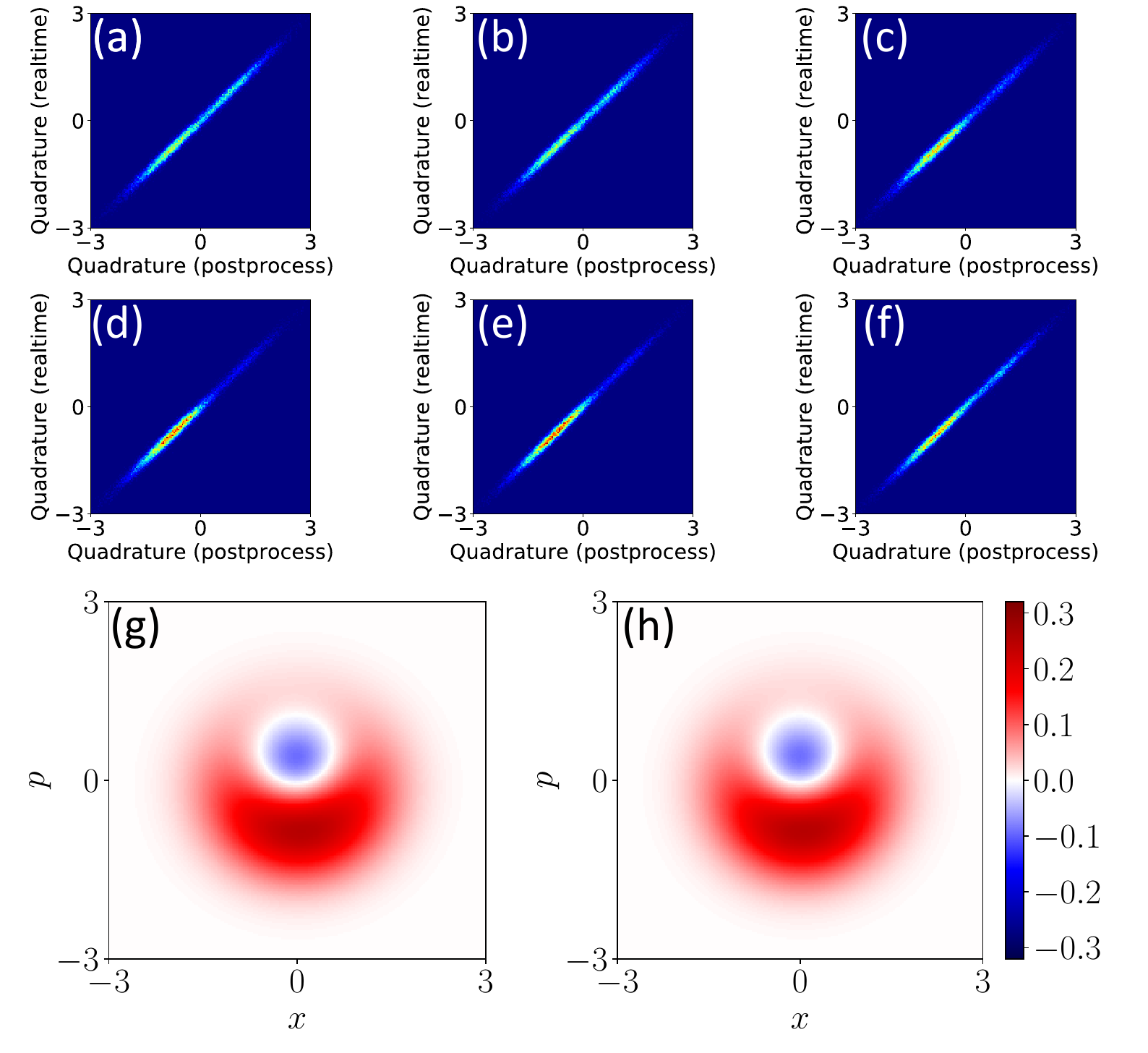}
\caption{Comparison between digital processing method and real-time method for the point C in Fig.\ \ref{fig:exp_result}. (a-f), Correlations between quadrature values obtained via digital postprocessing and the real-time method for the phase $0^\circ$, $30^\circ$, $60^\circ$, $90^\circ$, $120^\circ$, and $150^\circ$, respectively. (g) and (h), reconstructed Wigner functions using the quadrature values of the digital processing method and the real-time method, respectively.}\label{fig:postprocess_and_realtime}
\end{figure}

As an example, Fig.\ \ref{fig:postprocess_and_realtime} show the quadrature distribution and the reconstructed Wigner functions of the target state that is equivalent to the Wigner function C of Fig.\ \ref{fig:exp_result}, analyzed with both the digital postprocessing and the real-time measurements. We observe that for any phase, the quadrature values of both method are highly correlated with a correlation coefficient $>0.99$. The reconstructed Wigner functions are also qualitatively the same for both methods. For all the other superpositions, the situations are also similar and the correlation coefficients for all the generated states at all quadratures are $>0.98$. Therefore, we can conclude that we have succeeded in the real-time measurements of the superposition of the Fock states.

%%%%%%%Appendix C%%%%%%%%
\section*{APPENDIX C: Detailed evaluations of the generated states}
\label{sec:appendixC}
\begin{figure}[htp]
\centering
\includegraphics[width=\columnwidth]{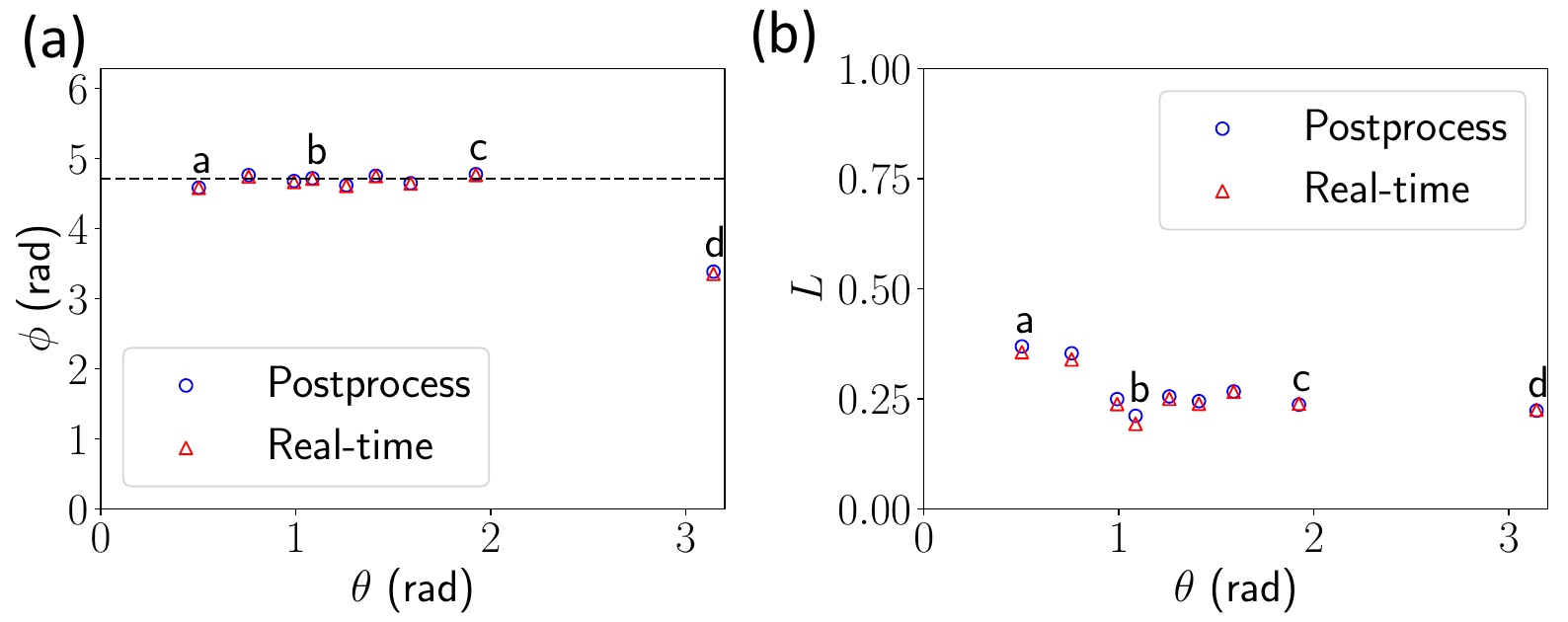}
\caption{Fitting of the $\phi$ (a) and $L$ (b) of the generated states. Blue circle markers: digital postprocessing method. Red triangle markers: real-time measurement method.}\label{fig:detailed_state}
\end{figure}
In the main text, we write down the parametrized superposition between the vacuum state and the single photon state as a pure state. However, the actual states are mixed states. If we let $L$ be the optical loss, the density matrix $\hat{\rho}(\theta,\phi,L)$ of the superposition can be written as
\begin{multline}
\hat{\rho}(\theta,\phi,L)=\\
\begin{pmatrix}
1-(1-L)\sin^2 \left( \frac{\theta}{2}\right) & \frac{1}{2}\sin(\theta)\mathrm{e}^{-i\phi} \sqrt{1-L} \\
\frac{1}{2}\sin(\theta)\mathrm{e}^{i\phi} \sqrt{1-L} & \sin^2 \left( \frac{\theta}{2}\right) \left( 1-L \right)
\end{pmatrix}  \label{eq:Superposition01}
\end{multline}
As $\theta$ is the parameter we mainly adjust, we will see whether the parameters $\phi$ and $L$ for each generated state are consistent over $\theta$ or not.

Figure \ref{fig:detailed_state} shows the fitting of $\phi$ and $L$ between the experimentally reconstructed density matrices and $\hat{\rho}(\theta,\phi,L)$ as a function of $\theta$. We observe that $\phi$ is closed to $3\pi/2$ within a range of $\pm0.2$ rad. The only exception here is the $\theta=\pi$, i.e. the single photon state which is a phase insensitive state, rendering the fitting inaccurate. On the other hand the values of $L$ are mostly gathered around $0.25$ for most of the $\theta$ except for the small $\theta$. This behavior is expected as small $\theta$ means that most of the contribution of the state is due to vacuum state which is barely affected by losses, making the fitting of $L$ unreliable at a small $\theta$.

\bibliography{exprise02bib}

\end{document}